\documentclass[%
 reprint,
 amsmath,amssymb,
 aps,
prd,
]{revtex4-1}

\usepackage{graphicx}
\usepackage{dcolumn}
\usepackage{bm}
\usepackage{subfigure}

\begin{document}


\title{Super-Chandrasekhar White Dwarfs with Magnetic-dependent Equation of State}


\author{Qi-Xiang Zou  and Xin-He Meng$^*$ \email[]{xhm@nankai.edu.cn}
\thanks{xhm@nankai.edu.cn}}
\affiliation{School of Physics, Nankai University, Tianjin 300071, China\\
State Key Laboratory of Theoretical Physics, ITP-CAS, Beijing, P.R.China}


\date{\today}

\begin{abstract}
Recently, some over-luminous Ia supernovaes are found, suggesting that their progenitors are white dwarfs more massive than the Chandrasekhar limit, which perhaps result from ultra-strong magnetic field inside the white dwarfs. We present an equation of state, explicitly magnetic-dependent and analytically practicable, and observe that the change of equation of state due to magnetic field waning along radius will so significantly influence the configuration of a white dwarf that its density does not monotonically decrease,  but goes down at first, re-peaks near the crust and falls off again. As a supernovae will, in single degenerate Ia supernovae system, leave the remnant of its companion and a neutron star (pulsar star), we point out that the observations of these objects can put our model into tests.
\end{abstract}


\maketitle

\section{\label{sec:level1}Introduction}
A main sequence star of low or mediate mass will finally evolve into a white dwarf (WD) after supernova explosion. S. Chandrasekhar\cite{Chandrasekhar:1931} has shown that a WD will not sustain a mass over 1.4 $M_{\bigodot } $, at which  very mass a WD, after absorbing mass from its companion, will explode again and become the Ia Supernovae (SNIa). This is the Single Degenerate (SD) mechanism of SNIa formation\cite{whelan1973binaries}\cite{Wang:2012za}, and suggests that SNIas are all of with nearly equal luminosity and hence can be regarded as  the cosmic standard  candle to measure cosmology distance, contributing to the surprise discovery of the accelerative expansion of our observable cosmos\cite{Riess:1998cb}, and revolutionizing our understanding of fundamental physics world even radically extending our conventional gravity and cosmology models. However, some SNIas, such  as SN 2007if, are recently found whose progenitors are even twice as massive as Chandrasekhar limit (ChL)\cite{Howell:2006vn}\cite{Hicken:2007ap}\cite{Yamanaka:2009dp}\cite{Scalzo:2010xd}\cite{Silverman:2010bh}\cite{Taubenberger:2010qv}, for which some authors try to introduce electric field\cite{Liu:2014jna} or rotation contributions\cite{Maeda:2008he} into a WD to bypass ChL within the SD SNIa model. Besides, Double Degenerate (DD) Mechanism\cite{tutukov1981evolutionary}, where SNIa is ignited by two WDs merging together, can account for a wide range of SNIa luminosity while fails to explain why most SNIas are similar.

In Chandrasekhar's original work\cite{Chandrasekhar:1931} , the equation of state (EOS) is determined by the degenerate pressure amongst relativistic Fermi gas of electrons. To accommodate the recently found super-massive DWs, the work by U. Das and B. Mukhopadhyay (D$\& $M) \cite{Das:2012ai}\cite{PhysRevLett.110.071102} issues that, since ultra-strong magnetic field (USM) will quantize free electron gas into Landau levels and soften the EOS, a WD will be 2.58$M_{\bigodot } $'s massive when inside is with a constant magnetic field as strong as $10^{18} Gauss (G)$. This work is, however, challenged for over-simplicity and on stability problems\cite{Coelho:2013bba}\cite{Chamel:2013tfa}\cite{PhysRevD.91.028301}. Not very long, U. Das and his collaborators\cite{Das:2014ssa} refine their previous work\cite{Das:2012ai} by relativistically considering a density dependent isotropic magnetic field, and response some critics\cite{Das:2013kga}\cite{PhysRevD.91.028302}. Meanwhile, R. Nityuananda and S. Konar\cite{Nityananda:2013yua} attempt to constrain the configuration of a magnetic WD but their result was soon proven erroneous\cite{PhysRevD.91.029904}.

The EOS used in the above papers is simply a quadratic approximation for only $\nu =1$ Landau Level in USM existence, because the original EOS with Landau levels is in the awkwardly complicated form of algebraic summation. As the magnetic field will decrease with radius, the EOS will surely shift accordingly, which is, however, overlooked. We will reduce the Landau level EOS to a feasibly analytical form without loss of accuracy, which will simplify the discrete summation and accurately demonstrate the influence of varying magnetic field on the multi-level EOS\cite{Das:2012ai}\cite{Suh:1999tg}.

General Relativity contains high non-linearity from complication of curvature, with appearance of metric tensor in energy-momentum tensor, and even interlocked couple of electromagnetism in the magnetized astrophysics. Especially, the structure equation  will be insurmountable to solve exactly when magnetic field destroys the simple spherical symmetry. In D$\& $M, they circumvent such complexity by ideally considering constant magnetic field in Newtonian frame.  Some authors also try to reduce the complexity by unnaturally introducing into a spherically symmetric metric with anisotropy in energy momentum tensor\cite{Bowers:1974}. Only few works\cite{Das:2014owa}\cite{Bocquet:1995je} deal with non-spherical cases by virtue of, for example, GRMHD code\cite{Bucciantini:2010ax}, but the magnetic fields in their works are much too weak.  It is well-know that as the gravitation field inside a white dwarf is weak\cite{Max} , the linear theory will be sufficient enough for describing. Hence, we are going to develop the general equation of equilibrium 
in the framework of linearized General Relativity, where anisotropy  property can be treated much easier.

We arrange this paper as follows. The next section will present the linearly relativistic equation of equilibrium for our working frame. The third section will be devoted to magnetic field distribution by modelling with two regions, and the forth section deals with the equation of state in the strong magnetic field. Finally, we will summarize the result and  draw the conclusion with some discussions for possible perspectives. Throughout this paper, the speed of light $c=1$ is taken, and the Minkowski metric is set to be $diag \left( {-1,1,1,1} \right)  $ as commonly adopted.
\section{Linearized General Relativity}
Denote the deviation from Minkowski metric as,
\begin{equation}
\gamma_{\mu\nu } = g_{\mu\nu } -\eta_{\mu\nu } ,
\end{equation}
so that the Einstein's field equation, expanded to the linear order and imposed the Lorentz gauge, will be
\begin{equation}
\square\gamma_{\mu\nu } =-16\pi G \left( T_{\mu\nu } -\frac{1}{2} T\cdot\eta_{\mu\nu }\right) .
\end{equation}

If inserting into $T_{\mu\nu } $ as summed from that of ideal fluid and electromagnetic field, noticing that $T=3p-\rho $, regarding $\frac{1}{2}\gamma_{00} $ as the potential $\phi $, we get the modified Poisson Equation,
\begin{equation} \label{eq:one}
\vartriangle\phi =4\pi G\left( \rho +\rho_B +3p\right)  , 
\end{equation}
where $\rho_B =\frac{1}{8\pi } B^2 $ is the energy density of magnetic field. We ignore the rest part of $\gamma_{\mu\nu } $ for they will not appear in the equilibrium equation.

The equation of motion can not be directly obtained from linear theory as $\nabla_{\mu } T^{\mu\nu } =0$; 
however, for ideal fluid in the frame of General Relativity, if we assume that there is no 4-current inside a star, the equilibrium equation is given as \cite{MTW}
\begin{equation}
\nabla p=-(\rho +p)\nabla \ln\sqrt{-g_{00} } ,
\end{equation}
which, expanded to the leading order, will be
\begin{equation}  \label{eq:two}
\nabla p=-(\rho +p)\nabla \phi  .
\end{equation}
These equations will be formed as a close system if blessed with the equation of state and magnetic distribution.

For illustrious purpose,  consider the spherically symmetric case, in which Eqs.~(\ref{eq:one}) and (\ref{eq:two}) will be reduced as,
\begin{eqnarray}
\frac{\mathrm{d} M}{\mathrm{d} r} &=& 4\pi r^2 \left( \rho +\rho_B +3p\right)\\
\frac{\mathrm{d} p}{\mathrm{d} r} &=& -(\rho +p)\frac{GM}{r^2 } 
\end{eqnarray}
\section{Magnetic Field}
The magnetic field inside an ultra-strongly magnetic WD (USMWD) is not clearly known. As the magnetic flux $BR^2 $ is preserved through a collapse process \cite{Ginzberg:1964}\cite{thorne1964resistance}\cite{woltjer1964x}, an usual DW can succeed from its progenitor a magnetic field at least 5 orders stronger. However, accretion might also play an indispensable role in the formation of an USMWD. Das and his collaborators\cite{Das:2012ai}\cite{PhysRevLett.110.071102} initially illustrate the effect of a constant field, which, for magnetic field will dwindle at least two orders from center to surface, are only a simplified toy model with some debattings.

However, a well-known magnetic field distribution is indeed constant in the center core while decreasing rapidly out of the core. As it is known that for an uniformly magnetized sphere in vacuum, magnetic field is constant inside the sphere and dipole distribution outside\cite{Jackson} ,
\begin{eqnarray}
\mathbf{B} =
\begin{cases}
\mathbf{B}_0 & r\leqslant r_0 \\
-\frac{1}{2r_0^3 }\nabla\left( \mathbf{B}_0\cdot\mathbf{r} / \mathbf{r}^3 \right) & r> r_0
\end{cases}
\end{eqnarray} 
where $r_0 $ is the radius of inner core.

Because the main purpose of this paper is to illustrate how the impact of decreasing magnetic field on EOS will affect the structure of an USMWD, we will assume that the magnetic field is spherically symmetric and, in light of the distribution of uniformly magnetic sphere, we take the magnetic field as constant inside an inner core in the center of a WD and inversely cubic outside, that is, by generalizing the above distribution directly
\begin{eqnarray}
B =
\begin{cases}
B_0 & r\leqslant r_0 \\
B_0\cdot r_0^3 /r^3  & r> r_0
\end{cases}
\end{eqnarray} 
where $r_0 $ is the radius of the inner core. The direction of magnetic field induction will accommodate to satisfy Maxwell's equation.

In addition, the conservation law of the magnetic flux also suggests that the ruminant neutron star of an SNIa  possess a magnetic field of at least $10^{20} G$. As neutron stars are much easier to detect, it is very likely to discover USM neutron star by the pulsar star properties like the timing. This will justify the USMWD hypothesis whether the neutron star is from a SNIa or not, because if it is, its progenitor's magnetic field will be sufficiently strong; if not, at least the existence of a main sequence star of magnetic field stronger than $10^{10} G$ will be confirmed, which also can be the progenitor of an USMWD.

\section{Equation of States}

Landau level describes a free electron moves in an uniform magnetic field B. Set the magnetic field along z direction, i.e., the vector potential be 
\begin{equation}
A_\mu =(0, By, 0, 0);
\end{equation}
so the Landau energy level, given by Dirac equation, is
\begin{equation}
E_{\nu ,p_z } =\sqrt{p^2_z +m_e (1+2\nu B_D )} 
\end{equation}
where $\nu =j+\frac {1}{2} +\sigma $ is the quantum number of Landau level, and $B_D =B\hbar e/m_e^2 $. So the maximium of $\nu $ is $\nu_m =(\epsilon_F^2 -1)/2B_D $ where the $\epsilon_F$ is the Fermi's level, and the EOS will be \cite{Das:2012ai} 
\begin{widetext}
\begin{eqnarray}
\rho &=& \frac{2\mu_e m_H B_D }{(2\pi )^2 \lambda_e^3 }\times\left(\sqrt{\epsilon_F^2 -1} +2\theta (\nu_m -1)\sum^{[\nu_m ] }_{\nu =1} \sqrt{\epsilon_F^2 -1-2\nu B_D }  \right) \\
p &=& \frac{2m_e B_D }{(2\pi )^2 \lambda_e^3 }\times \left(\eta (\sqrt{\epsilon_F^2 -1} ) +2\theta (\nu_m -1)\sum^{[\nu_m ]}_{\nu =1} (1+2\nu B_D )\eta (\sqrt{\frac{\epsilon_F^2 -1-2\nu B_D }{1+2\nu B_D }} ) \right)
\end{eqnarray}
\end{widetext}
where $\mu_e $ means molecular weight per electron, $\lambda_e $ Compton wavelength of electron, $m_H $ mass of Hydrogen atom, $\epsilon_F $ Fermi energy, $\theta (z)$ is the step function that is equal to constant one except vanishing when $z$ negative, and
\begin{equation}
\eta (z)=\frac{1}{2}\left[ z\sqrt{1+z^2 } -\ln (z+\sqrt{1+z^2 } )\right]
\end{equation}
\begin{figure} \centering    
\subfigure[$S_\rho $ and its deviation] { \label{fig1:a}     
\includegraphics[width=0.9\columnwidth]{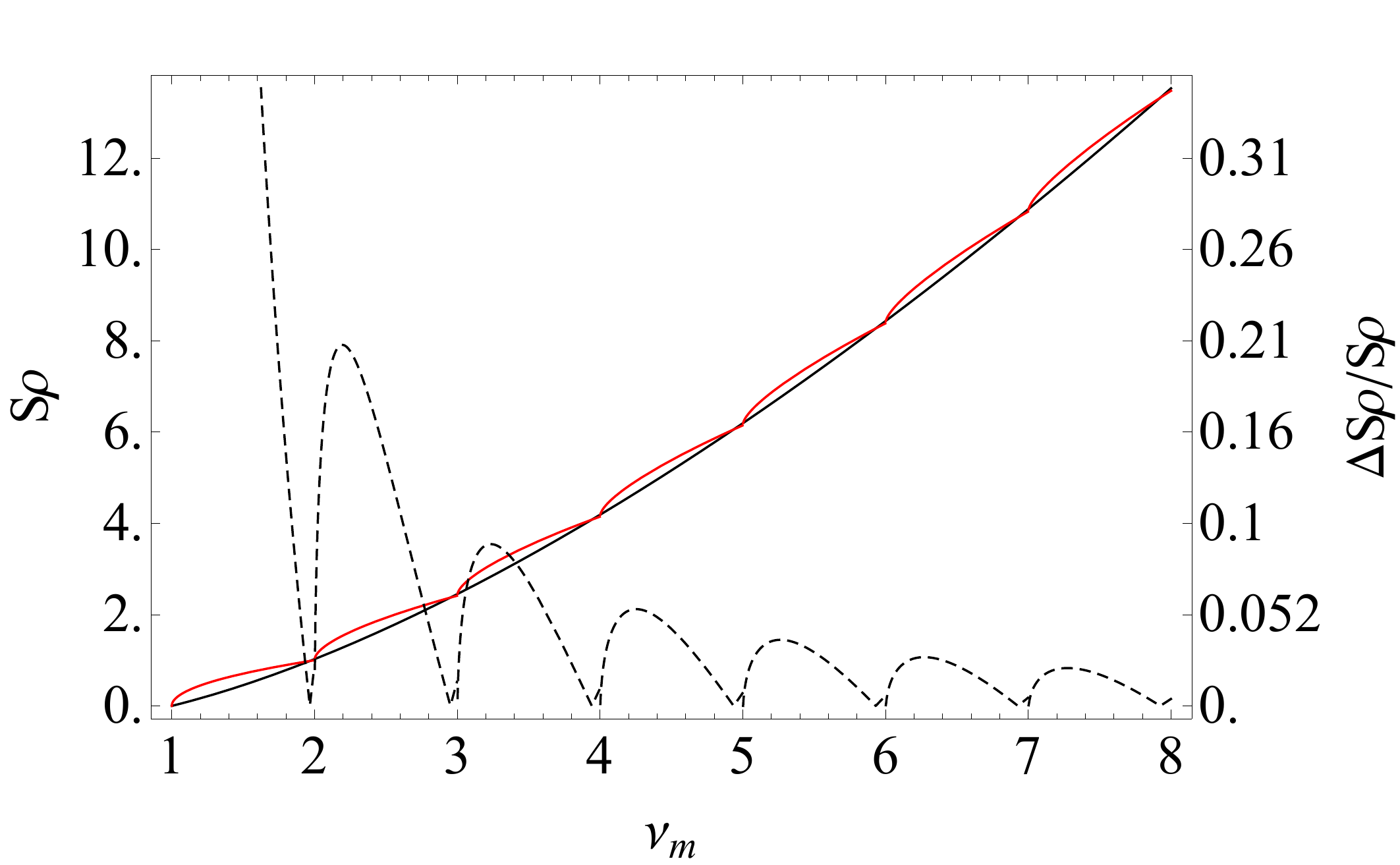} 
}   
\subfigure[$S_p $ and its deviation] { \label{fig1:b}     
\includegraphics[width=0.9\columnwidth]{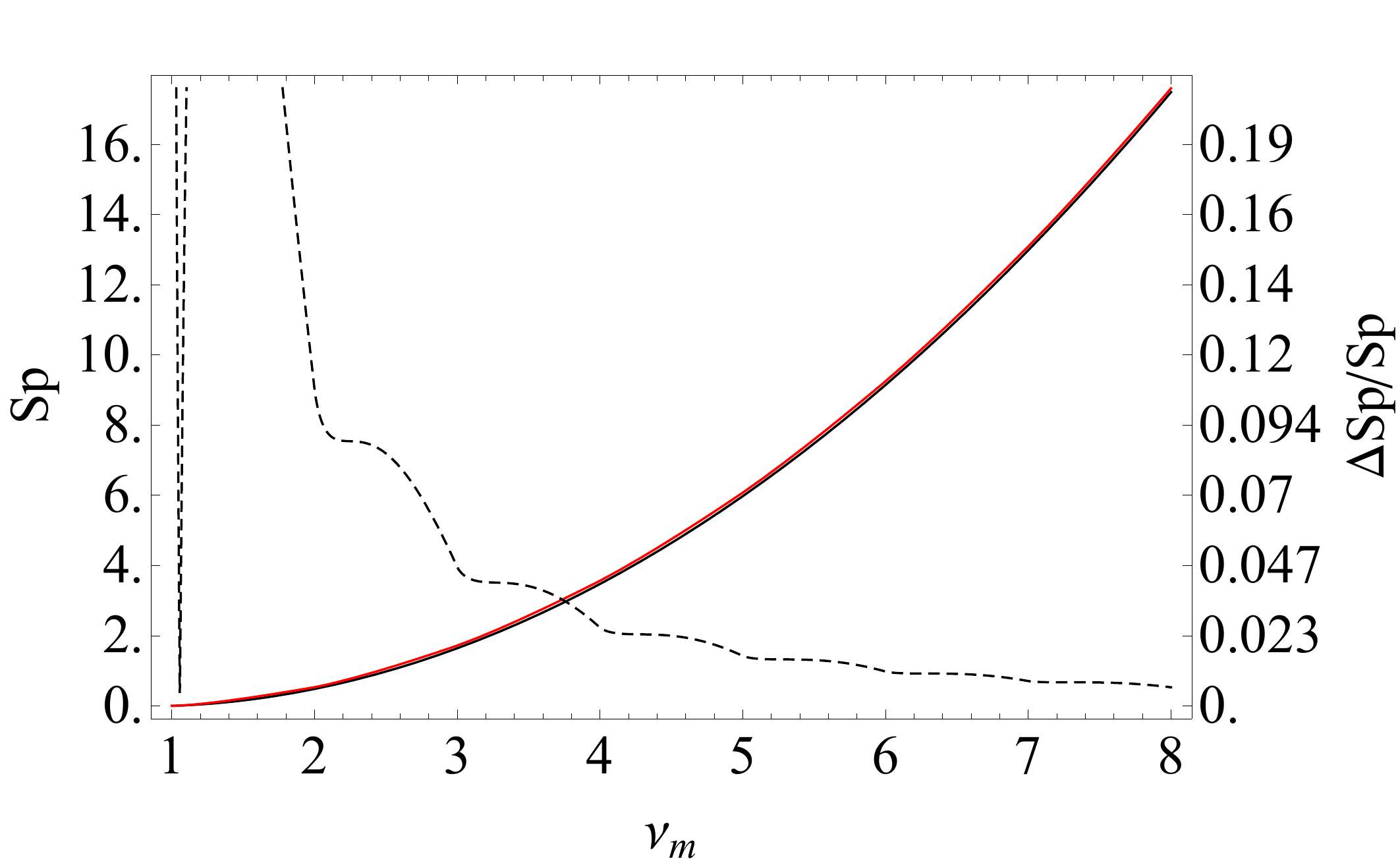}     
}     
\caption{Exactitude of $S_\rho $ and $S_p $. The black lines represent $S_\rho $ and $S_p $ and the red lines stand for the accurate original discrete summation. We can see that they are very close and get closer and closer as $\nu_m $ increases. The relative errors are plotted in dashed gray lines.}     
\label{fig1}     
\end{figure}
The summation terms that appear in EOS cause awkwardness and inconvenience for practical use, but it can be surmounted by the famous Eular's formula \cite{SpeFunc},
\begin{equation}
\begin{split}
\sum_{i=1}^{m-1} F(i)= &\int_0^m F(x)\mathrm{d} x-\frac{1}{2} (F(0)+F(m) \\
&-\sum_{k=a}^{n}\frac{(-1)^k B_k }{(2k)!}\cdot [F^{(2k-1} (m)-F^{(2k-1} (0)] \\
&+\xi\frac{(-1)^{k+1} B_{k+1} }{(2k+2)!}\cdot [F^{(2k+1} (m)-F^{(2k+1} (0)] 
\end{split}
\end{equation}
where $B_k $ stands for the $k$th Bernoulli number, as 1/6, 1/30, 1/42, etc. and $\xi $ some number between 0 and 1.  For $ \epsilon_F \gg 1$ and $B_D\gg 1$, the Landau EOS can be (for details, see Appendix A)
\begin{eqnarray}
\rho &=& \frac{2\mu_e m_H B_D }{(2\pi )^2 \lambda_e^3 }\cdot\left[ \sqrt{2B_D\nu_m }+\theta (\nu_m -1)S_{\rho } (\nu_m )\right] \\
p &=& \frac{2m_e B_D^2 }{(2\pi )^2 \lambda_e^3 }\left[ \nu_m +2\theta (\nu_m -1)S_p (\nu_m )\right]
\end{eqnarray}

$\nu_m $ appears in the above EOS as a continuous parameter because it sufficiently approximates the original discrete summation as illustrated in FIG.~\ref{fig1} .

\section{Results}
\begin{figure}[b]
\includegraphics[scale=0.4]{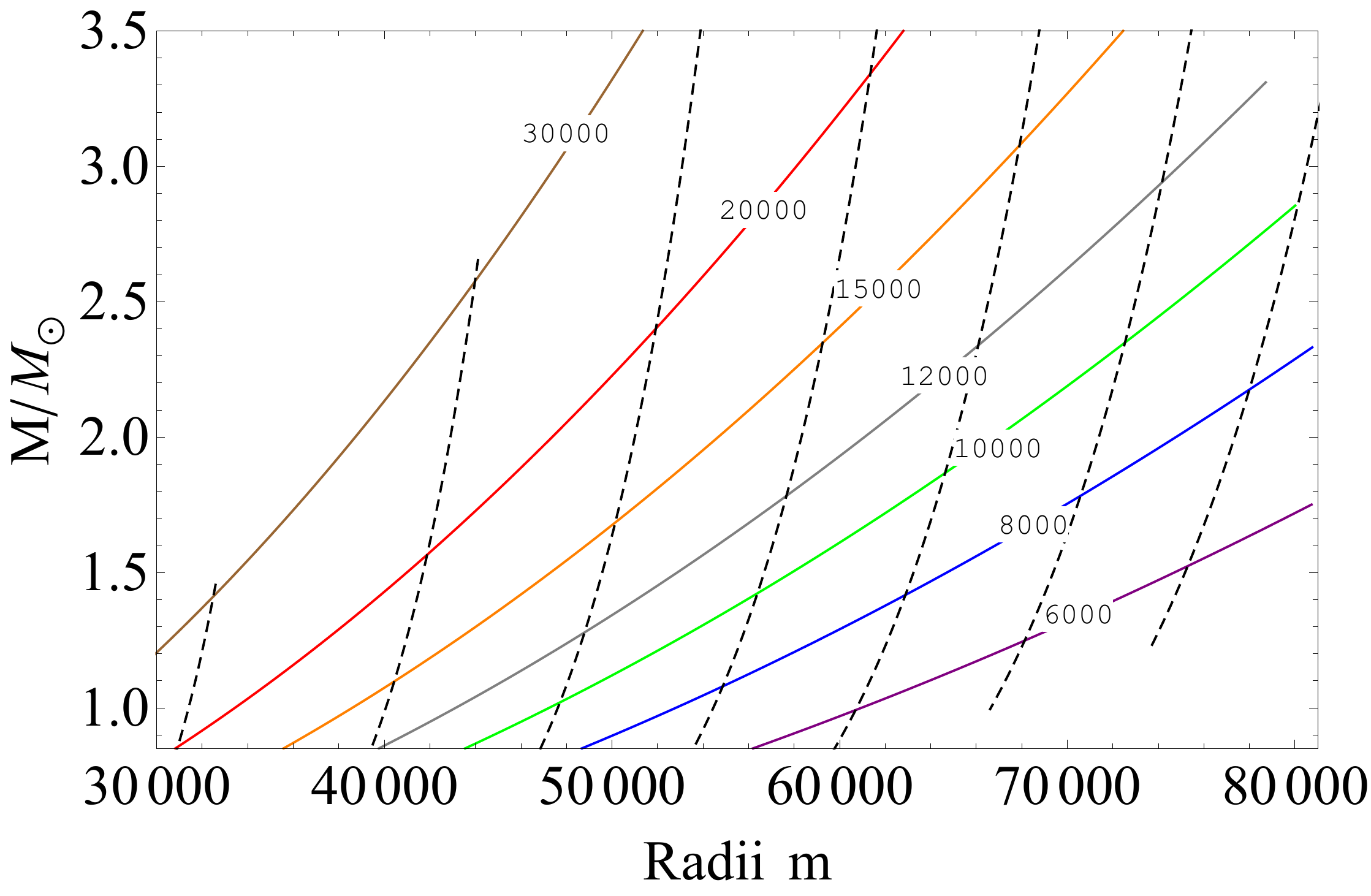} 
\caption{\label{fig2} Mass-Radii Relations.  The M-R relation of $B_D =30000, 20000, 15000, 12000, 10000, 8000, 6000 $ is drawn in brown, red, orange, gray, green, blue and purple. The dashed black lines in the figure represent the M-R relation for same inner core radii, from 500m to 2000m at the interval of 250m.}
\end{figure}

Given the EOS, Magnetic field distribution, and  structure equations, we obtain the Mass-Radii relations as plotted in FIG.~\ref{fig2}. Masses will increase with respect to core radii and central strength of magnetic field, while Radii are almost determined by the $r_0 $ and independent of $B_D $. 
Despite the magnetic field, $\rho $, $p$, and $\nu_m $ are also nearly uniform in the inner core, as for $\nu_m $ inside, $p\ll\rho $, $\rho_B \ll\rho $, for its derivative,
\begin{equation}
\nu_m '\approx\frac{\sqrt{2}\mu_2 m_H }{2m_e B_0^{1/2} }\cdot\frac{GM}{r^2 } \sim 3.3\times 10^{-14} B_0 r.
\end{equation} 
However, outside the inner core, the derivative of $B_D $ will also contribute to $\nu_m ' $, together with the accumulation of mass, so that $\nu_m $ will increase significantly and $p$ will begin dropping, i.e., $p$ keeps constant inside whereas decreases outside the inner core, and the later $p$ dwindles, the slower $p$ vanishes; hence the increasing of radii and mass with  respect to $r_0 $. The influence of magnetic field strength can be understood as follows. M, as is dominated by $\rho $, is proportional to at most $B_D^{3/2} $ while $p\sim B_D^2 $; therefore, $p$ will increase faster than its derivative.
 
\begin{figure} \centering    
\subfigure[ $\rho -r$ relation] { \label{fig3:a}     
\includegraphics[width=0.9\columnwidth]{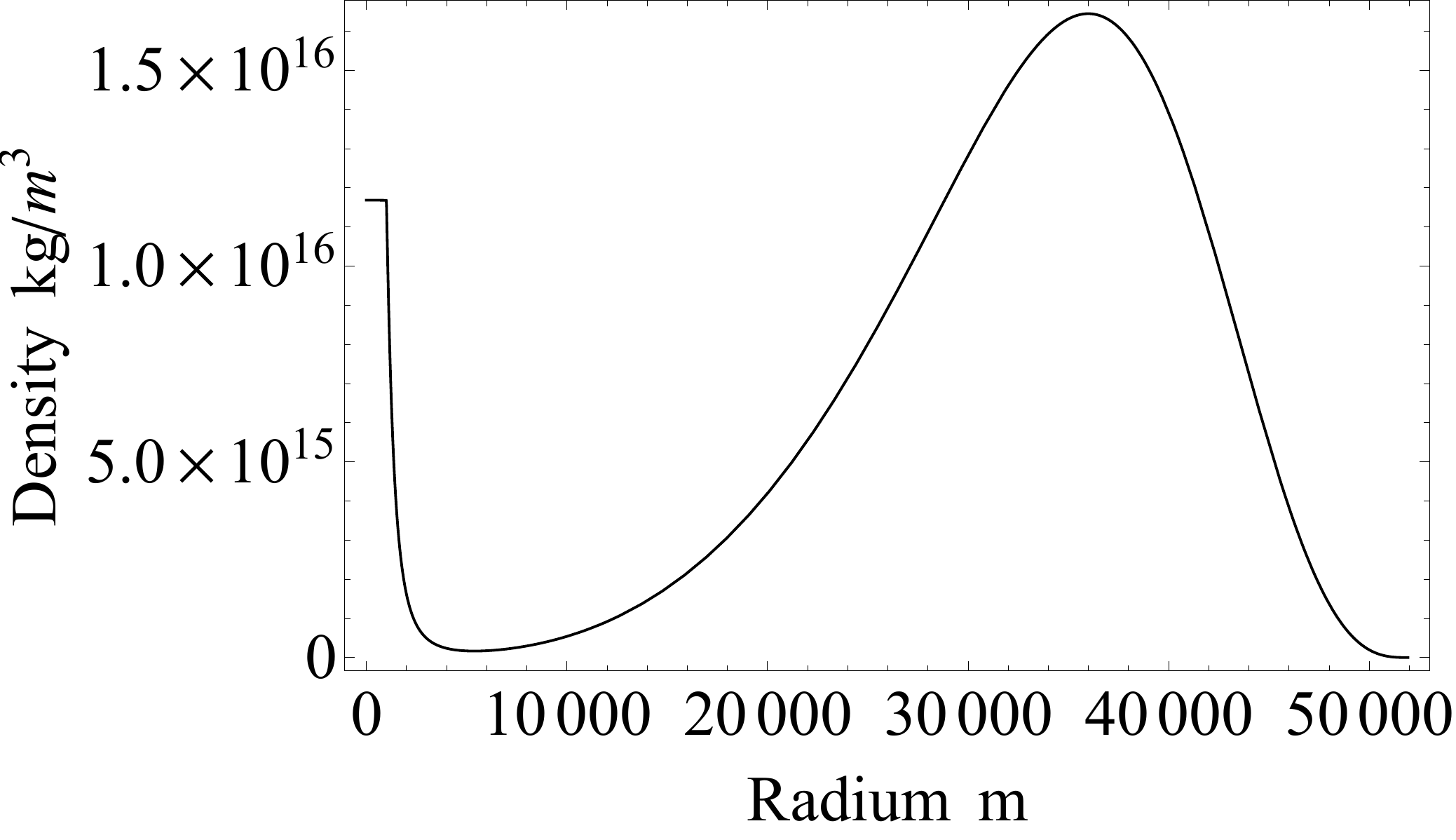} 
}   
\subfigure[ $\nu_m -r$ relation] { \label{fig3:b}     
\includegraphics[width=0.9\columnwidth]{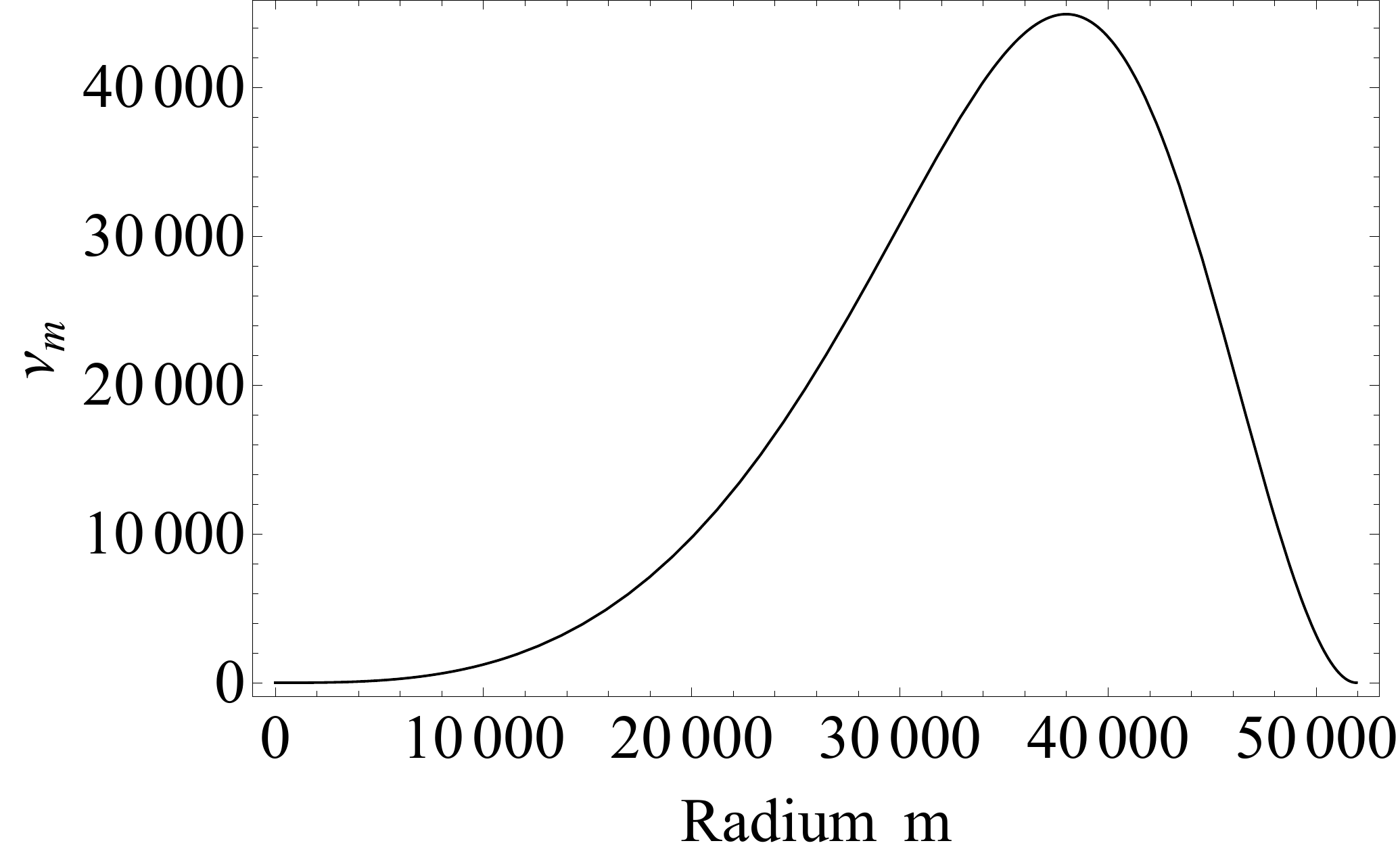}     
}     
\caption{Structure of a WF with $B_D =20000$, and $R0=1000m$ }     
\label{fig3}     
\end{figure}

We will study a typical USMWD with $B_D =20000$, $R_0 =1000m$, whose mass is solved to be $2.40M_{\bigodot } $ and radius 51.930 km. Its density and $\nu_m $ distribution is presented as in FIG.~\ref{fig3} . FIG.~\ref{fig3:a} indicates that $\rho $ does not monotonously decrease with respect to r. In the contrary it will peak again near the crust surface, which violates P. Bara's negative result on the magnetic influence structure \cite{Bera:2014wja}. When $\nu_m $ increases as $B_D $ goes down, $p\sim\rho^{4/3} B_D^{1/3} $. Accordingly, $\rho $ will increase if $p$ decreases slower than $1/r$. Such structure undermines the Global stability as mentioned in Ref.~\cite{Chamel:2013tfa} , 
\begin{equation}
\frac{E_{mag} }{E_{grav} } =\frac{\int_0^M \frac{\rho_B }{\rho }\mathrm{d} m}{\frac{1}{2} \int_0^M \Phi\mathrm{d} m} .
\end{equation}
In this expression, $E_{mag} \approx \int 4\pi r^2 \rho_B \mathrm{d} r$ is nearly independent of $\rho (r)$ whereas $E_{grav} =\int M(r)M'(r)/r \mathrm{d} r$ peaks at $M_0 \delta (r) $ and wanes when $\rho $ is not centralized near the center. The global instability for this DW is as high as 60.2.

In addition, as magnetic field  generally introduces dipole effect, the dipole momentum of an USMWD will be  amplified. If an USMWD is found in a binary system, the "re-peaking" effect can be put to test by comparing the dipole mass momentum measured from the shift of the binary system period with both its mass and radii.
\section{Conclusion}
We have derived the analytical form of Landau level EOS, presented
 the M-R relation for USMWD, and have found the "re-peaking" effect of  the density profile for the EOS becoming harder when magnetic field wanes. We also prove that USMWD's high magnetic field hypothesis suggests that the existence of highly magnetic neutron star and of SNIa ruminant companion. If they are both discovered, USM will be possible and SD SNIa model be confirmed. At least, we can conclude that the progenitors of some over-luminous SNIas will indeed be USMWD. We should say that the concrete mechanism and distribution of such USM are still unclear, which are however, indispensable for deriving the statistical distribution of USMWDs. This is only the first step for our work, though only touching some aspects of the interesting star forming processes involved and with upcoming more accurate astrophysics observational data we are keeping on studying the accretion process in order to understand further the origin of the USM, as it may shed new light on our novel understanding the extrema conditional physics.
\begin{acknowledgments}
The authors appreciate Professor Saibal Ray for interesting discussions for a long time communications on related topics. This work is partially 
funded by the Natural Science Foundation of China (NSFC) and the State Key Laboratory of Theoretical Physics, ITP-CAS of China.
\end{acknowledgments}
\appendix
\section{$S_\rho $ and $S_p $}
In this appendix we will present more details for the analytical treatment to the complicated EOS.
We take Eular's formula to the order $n=1$ and define $\Sigma_E $ as,
\begin{equation}
\begin{split}
\Sigma_E (&F(i),l+1,m-1) \\
&= \int_{l}^m F(x)\mathrm{d} x-\frac{1}{2} [F(l)+F(m) ]+\frac{1}{12} [F' (m)-F' (l)]
\end{split}
\end{equation}
So $S_\rho $,
\begin{equation}
\begin{split}
S_\rho^* &\approx 0+1+\Sigma_E (\sqrt{\nu_m -i},1,\nu_m -2) \\
&=\frac{-1-5\sqrt{\nu_m } -12\nu_m +16\nu_m^2 }{24\sqrt{\nu_m } }
\end{split}
\end{equation}
$S_\rho =S_\rho^* +\frac{1}{12} $ to secure that $S_\rho $ vanishes at $\nu_m =1$. Similarly, for $S_p $, 
\begin{equation}
\begin{split}
S_p^* &= \sum_{i=1}^{\nu_m } i\cdot\left[\sqrt{-1+\frac{\nu_m }{i} }\sqrt{\frac{\nu_m }{i} } -\ln (\sqrt{-1+\frac{\nu_m }{i} } +\sqrt{\frac{\nu_m }{i} } )\right] \\
&\approx \sum_{i=1}^{\nu_m } (p1(i) +p2(i) ) \\
&\approx \Sigma_E (p1,1,\nu_m -1)+p1(\nu_m ) \\
&\;\;\,\;\,\;\, +\Sigma_E (p2,2,\nu_m -1)+p2(1)+p2(\nu_m )\\
&= \frac{1}{168}\left[ -84\nu_m +(315-372\ln 2) \nu_m^2 -7\ln \frac{\nu_m }{2} \right]
\end{split}
\end{equation}
where $p1(i)$ and $p2(i)$ are respectively as
\begin{eqnarray}
p1(i) &=& 2\nu_m\frac{i-\nu_m }{i-2\nu_m } +\frac{i^{\frac{5}{2} }\sqrt{\nu_m }\ln 2}{\nu_m^2 } -\frac{i}{2} \ln 4\nu_m  \\
p2(i) &=& \frac{i}{2}\ln i
\end{eqnarray}
Similarly, we can  manipulate $S_p =\frac{49}{50} (S_p^* -S_p^* \vert_{\nu_m =1} )$.

\bibliography{wd}

\begin{thebibliography}{36}%
\makeatletter
\providecommand \@ifxundefined [1]{%
 \@ifx{#1\undefined}
}%
\providecommand \@ifnum [1]{%
 \ifnum #1\expandafter \@firstoftwo
 \else \expandafter \@secondoftwo
 \fi
}%
\providecommand \@ifx [1]{%
 \ifx #1\expandafter \@firstoftwo
 \else \expandafter \@secondoftwo
 \fi
}%
\providecommand \natexlab [1]{#1}%
\providecommand \enquote  [1]{``#1''}%
\providecommand \bibnamefont  [1]{#1}%
\providecommand \bibfnamefont [1]{#1}%
\providecommand \citenamefont [1]{#1}%
\providecommand \href@noop [0]{\@secondoftwo}%
\providecommand \href [0]{\begingroup \@sanitize@url \@href}%
\providecommand \@href[1]{\@@startlink{#1}\@@href}%
\providecommand \@@href[1]{\endgroup#1\@@endlink}%
\providecommand \@sanitize@url [0]{\catcode `\\12\catcode `\$12\catcode
  `\&12\catcode `\#12\catcode `\^12\catcode `\_12\catcode `\%12\relax}%
\providecommand \@@startlink[1]{}%
\providecommand \@@endlink[0]{}%
\providecommand \url  [0]{\begingroup\@sanitize@url \@url }%
\providecommand \@url [1]{\endgroup\@href {#1}{\urlprefix }}%
\providecommand \urlprefix  [0]{URL }%
\providecommand \Eprint [0]{\href }%
\providecommand \doibase [0]{http://dx.doi.org/}%
\providecommand \selectlanguage [0]{\@gobble}%
\providecommand \bibinfo  [0]{\@secondoftwo}%
\providecommand \bibfield  [0]{\@secondoftwo}%
\providecommand \translation [1]{[#1]}%
\providecommand \BibitemOpen [0]{}%
\providecommand \bibitemStop [0]{}%
\providecommand \bibitemNoStop [0]{.\EOS\space}%
\providecommand \EOS [0]{\spacefactor3000\relax}%
\providecommand \BibitemShut  [1]{\csname bibitem#1\endcsname}%
\let\auto@bib@innerbib\@empty
\bibitem [{\citenamefont {Chandrasekhar}(1931)}]{Chandrasekhar:1931}%
  \BibitemOpen
  \bibfield  {author} {\bibinfo {author} {\bibfnamefont {S.}~\bibnamefont
  {Chandrasekhar}},\ }\href {\doibase 10.1086/143324} {\bibfield  {journal}
  {\bibinfo  {journal} {Astrophys.J.}\ }\textbf {\bibinfo {volume} {74}},\
  \bibinfo {pages} {81} (\bibinfo {year} {1931})}\BibitemShut {NoStop}%
\bibitem [{\citenamefont {Whelan}\ and\ \citenamefont
  {Iben~Jr}(1973)}]{whelan1973binaries}%
  \BibitemOpen
  \bibfield  {author} {\bibinfo {author} {\bibfnamefont {J.}~\bibnamefont
  {Whelan}}\ and\ \bibinfo {author} {\bibfnamefont {I.}~\bibnamefont
  {Iben~Jr}},\ }\href@noop {} {\bibfield  {journal} {\bibinfo  {journal} {The
  Astrophysical Journal}\ }\textbf {\bibinfo {volume} {186}},\ \bibinfo {pages}
  {1007} (\bibinfo {year} {1973})}\BibitemShut {NoStop}%
\bibitem [{\citenamefont {Wang}\ and\ \citenamefont {Han}(2012)}]{Wang:2012za}%
  \BibitemOpen
  \bibfield  {author} {\bibinfo {author} {\bibfnamefont {B.}~\bibnamefont
  {Wang}}\ and\ \bibinfo {author} {\bibfnamefont {Z.}~\bibnamefont {Han}},\
  }\href {\doibase 10.1016/j.newar.2012.04.001} {\bibfield  {journal} {\bibinfo
   {journal} {New Astron.Rev.}\ }\textbf {\bibinfo {volume} {56}},\ \bibinfo
  {pages} {122} (\bibinfo {year} {2012})},\ \Eprint
  {http://arxiv.org/abs/1204.1155} {arXiv:1204.1155 [astro-ph.SR]} \BibitemShut
  {NoStop}%
\bibitem [{\citenamefont {Riess}\ \emph {et~al.}(1998)\citenamefont {Riess}
  \emph {et~al.}}]{Riess:1998cb}%
  \BibitemOpen
  \bibfield  {author} {\bibinfo {author} {\bibfnamefont {A.~G.}\ \bibnamefont
  {Riess}} \emph {et~al.} (\bibinfo {collaboration} {Supernova Search Team}),\
  }\href {\doibase 10.1086/300499} {\bibfield  {journal} {\bibinfo  {journal}
  {Astron.J.}\ }\textbf {\bibinfo {volume} {116}},\ \bibinfo {pages} {1009}
  (\bibinfo {year} {1998})},\ \Eprint {http://arxiv.org/abs/astro-ph/9805201}
  {arXiv:astro-ph/9805201 [astro-ph]} \BibitemShut {NoStop}%
\bibitem [{\citenamefont {Howell}\ \emph {et~al.}(2006)\citenamefont {Howell}
  \emph {et~al.}}]{Howell:2006vn}%
  \BibitemOpen
  \bibfield  {author} {\bibinfo {author} {\bibfnamefont {D.~A.}\ \bibnamefont
  {Howell}} \emph {et~al.} (\bibinfo {collaboration} {SNLS Collaboration}),\
  }\href {\doibase 10.1038/nature05103} {\bibfield  {journal} {\bibinfo
  {journal} {Nature}\ }\textbf {\bibinfo {volume} {443}},\ \bibinfo {pages}
  {308} (\bibinfo {year} {2006})},\ \Eprint
  {http://arxiv.org/abs/astro-ph/0609616} {arXiv:astro-ph/0609616 [astro-ph]}
  \BibitemShut {NoStop}%
\bibitem [{\citenamefont {Hicken}\ \emph {et~al.}(2007)\citenamefont {Hicken},
  \citenamefont {Garnavich}, \citenamefont {Prieto}, \citenamefont {Blondin},
  \citenamefont {DePoy} \emph {et~al.}}]{Hicken:2007ap}%
  \BibitemOpen
  \bibfield  {author} {\bibinfo {author} {\bibfnamefont {M.}~\bibnamefont
  {Hicken}}, \bibinfo {author} {\bibfnamefont {P.}~\bibnamefont {Garnavich}},
  \bibinfo {author} {\bibfnamefont {J.}~\bibnamefont {Prieto}}, \bibinfo
  {author} {\bibfnamefont {S.}~\bibnamefont {Blondin}}, \bibinfo {author}
  {\bibfnamefont {D.}~\bibnamefont {DePoy}},  \emph {et~al.},\ }\href {\doibase
  10.1086/523301} {\bibfield  {journal} {\bibinfo  {journal} {Astrophys.J.}\
  }\textbf {\bibinfo {volume} {669}},\ \bibinfo {pages} {L17} (\bibinfo {year}
  {2007})},\ \Eprint {http://arxiv.org/abs/0709.1501} {arXiv:0709.1501
  [astro-ph]} \BibitemShut {NoStop}%
\bibitem [{\citenamefont {Yamanaka}\ \emph {et~al.}(2009)\citenamefont
  {Yamanaka}, \citenamefont {Kawabata}, \citenamefont {Kinugasa}, \citenamefont
  {Tanaka}, \citenamefont {Imada} \emph {et~al.}}]{Yamanaka:2009dp}%
  \BibitemOpen
  \bibfield  {author} {\bibinfo {author} {\bibfnamefont {M.}~\bibnamefont
  {Yamanaka}}, \bibinfo {author} {\bibfnamefont {K.}~\bibnamefont {Kawabata}},
  \bibinfo {author} {\bibfnamefont {K.}~\bibnamefont {Kinugasa}}, \bibinfo
  {author} {\bibfnamefont {M.}~\bibnamefont {Tanaka}}, \bibinfo {author}
  {\bibfnamefont {A.}~\bibnamefont {Imada}},  \emph {et~al.},\ }\href {\doibase
  10.1088/0004-637X/707/2/L118} {\bibfield  {journal} {\bibinfo  {journal}
  {Astrophys.J.}\ }\textbf {\bibinfo {volume} {707}},\ \bibinfo {pages} {L118}
  (\bibinfo {year} {2009})},\ \Eprint {http://arxiv.org/abs/0908.2059}
  {arXiv:0908.2059 [astro-ph.HE]} \BibitemShut {NoStop}%
\bibitem [{\citenamefont {Scalzo}\ \emph {et~al.}(2010)\citenamefont {Scalzo},
  \citenamefont {Aldering}, \citenamefont {Antilogus}, \citenamefont {Aragon},
  \citenamefont {Bailey} \emph {et~al.}}]{Scalzo:2010xd}%
  \BibitemOpen
  \bibfield  {author} {\bibinfo {author} {\bibfnamefont {R.}~\bibnamefont
  {Scalzo}}, \bibinfo {author} {\bibfnamefont {G.}~\bibnamefont {Aldering}},
  \bibinfo {author} {\bibfnamefont {P.}~\bibnamefont {Antilogus}}, \bibinfo
  {author} {\bibfnamefont {C.}~\bibnamefont {Aragon}}, \bibinfo {author}
  {\bibfnamefont {S.}~\bibnamefont {Bailey}},  \emph {et~al.},\ }\href
  {\doibase 10.1088/0004-637X/713/2/1073} {\bibfield  {journal} {\bibinfo
  {journal} {Astrophys.J.}\ }\textbf {\bibinfo {volume} {713}},\ \bibinfo
  {pages} {1073} (\bibinfo {year} {2010})},\ \Eprint
  {http://arxiv.org/abs/1003.2217} {arXiv:1003.2217 [astro-ph.CO]} \BibitemShut
  {NoStop}%
\bibitem [{\citenamefont {Silverman}\ \emph {et~al.}(2011)\citenamefont
  {Silverman}, \citenamefont {Ganeshalingam}, \citenamefont {Li}, \citenamefont
  {Filippenko}, \citenamefont {Miller} \emph {et~al.}}]{Silverman:2010bh}%
  \BibitemOpen
  \bibfield  {author} {\bibinfo {author} {\bibfnamefont {J.~M.}\ \bibnamefont
  {Silverman}}, \bibinfo {author} {\bibfnamefont {M.}~\bibnamefont
  {Ganeshalingam}}, \bibinfo {author} {\bibfnamefont {W.}~\bibnamefont {Li}},
  \bibinfo {author} {\bibfnamefont {A.~V.}\ \bibnamefont {Filippenko}},
  \bibinfo {author} {\bibfnamefont {A.~A.}\ \bibnamefont {Miller}},  \emph
  {et~al.},\ }\href {\doibase 10.1111/j.1365-2966.2010.17474.x} {\bibfield
  {journal} {\bibinfo  {journal} {Mon.Not.Roy.Astron.Soc.}\ }\textbf {\bibinfo
  {volume} {410}},\ \bibinfo {pages} {585} (\bibinfo {year} {2011})},\ \Eprint
  {http://arxiv.org/abs/1003.2417} {arXiv:1003.2417 [astro-ph.HE]} \BibitemShut
  {NoStop}%
\bibitem [{\citenamefont {Taubenberger}\ \emph {et~al.}(2011)\citenamefont
  {Taubenberger}, \citenamefont {Benetti}, \citenamefont {Childress},
  \citenamefont {Pakmor}, \citenamefont {Hachinger} \emph
  {et~al.}}]{Taubenberger:2010qv}%
  \BibitemOpen
  \bibfield  {author} {\bibinfo {author} {\bibfnamefont {S.}~\bibnamefont
  {Taubenberger}}, \bibinfo {author} {\bibfnamefont {S.}~\bibnamefont
  {Benetti}}, \bibinfo {author} {\bibfnamefont {M.}~\bibnamefont {Childress}},
  \bibinfo {author} {\bibfnamefont {R.}~\bibnamefont {Pakmor}}, \bibinfo
  {author} {\bibfnamefont {S.}~\bibnamefont {Hachinger}},  \emph {et~al.},\
  }\href {\doibase 10.1111/j.1365-2966.2010.18107.x} {\bibfield  {journal}
  {\bibinfo  {journal} {Mon.Not.Roy.Astron.Soc.}\ }\textbf {\bibinfo {volume}
  {412}},\ \bibinfo {pages} {2735} (\bibinfo {year} {2011})},\ \Eprint
  {http://arxiv.org/abs/1011.5665} {arXiv:1011.5665 [astro-ph.SR]} \BibitemShut
  {NoStop}%
\bibitem [{\citenamefont {Liu}\ \emph {et~al.}(2014)\citenamefont {Liu},
  \citenamefont {Zhang},\ and\ \citenamefont {Wen}}]{Liu:2014jna}%
  \BibitemOpen
  \bibfield  {author} {\bibinfo {author} {\bibfnamefont {H.}~\bibnamefont
  {Liu}}, \bibinfo {author} {\bibfnamefont {X.}~\bibnamefont {Zhang}}, \ and\
  \bibinfo {author} {\bibfnamefont {D.}~\bibnamefont {Wen}},\ }\href {\doibase
  10.1103/PhysRevD.89.104043} {\bibfield  {journal} {\bibinfo  {journal}
  {Phys.Rev.}\ }\textbf {\bibinfo {volume} {D89}},\ \bibinfo {pages} {104043}
  (\bibinfo {year} {2014})},\ \Eprint {http://arxiv.org/abs/1405.3774}
  {arXiv:1405.3774 [gr-qc]} \BibitemShut {NoStop}%
\bibitem [{\citenamefont {Maeda}\ and\ \citenamefont
  {Iwamoto}(2009)}]{Maeda:2008he}%
  \BibitemOpen
  \bibfield  {author} {\bibinfo {author} {\bibfnamefont {K.}~\bibnamefont
  {Maeda}}\ and\ \bibinfo {author} {\bibfnamefont {K.}~\bibnamefont
  {Iwamoto}},\ }\href {\doibase 10.1111/j.1365-2966.2008.14179.x} {\bibfield
  {journal} {\bibinfo  {journal} {Mon.Not.Roy.Astron.Soc.}\ }\textbf {\bibinfo
  {volume} {394}},\ \bibinfo {pages} {239} (\bibinfo {year} {2009})},\ \Eprint
  {http://arxiv.org/abs/0811.2095} {arXiv:0811.2095 [astro-ph]} \BibitemShut
  {NoStop}%
\bibitem [{\citenamefont {Tutukov}\ and\ \citenamefont
  {Yungelson}(1981)}]{tutukov1981evolutionary}%
  \BibitemOpen
  \bibfield  {author} {\bibinfo {author} {\bibfnamefont {A.}~\bibnamefont
  {Tutukov}}\ and\ \bibinfo {author} {\bibfnamefont {L.}~\bibnamefont
  {Yungelson}},\ }\href@noop {} {\bibfield  {journal} {\bibinfo  {journal}
  {Nauchnye Informatsii}\ }\textbf {\bibinfo {volume} {49}},\ \bibinfo {pages}
  {3} (\bibinfo {year} {1981})}\BibitemShut {NoStop}%
\bibitem [{\citenamefont {Das}\ and\ \citenamefont
  {Mukhopadhyay}(2012)}]{Das:2012ai}%
  \BibitemOpen
  \bibfield  {author} {\bibinfo {author} {\bibfnamefont {U.}~\bibnamefont
  {Das}}\ and\ \bibinfo {author} {\bibfnamefont {B.}~\bibnamefont
  {Mukhopadhyay}},\ }\href {\doibase 10.1103/PhysRevD.86.042001} {\bibfield
  {journal} {\bibinfo  {journal} {Phys.Rev.}\ }\textbf {\bibinfo {volume}
  {D86}},\ \bibinfo {pages} {042001} (\bibinfo {year} {2012})},\ \Eprint
  {http://arxiv.org/abs/1204.1262} {arXiv:1204.1262 [astro-ph.HE]} \BibitemShut
  {NoStop}%
\bibitem [{\citenamefont {Das}\ and\ \citenamefont
  {Mukhopadhyay}(2013)}]{PhysRevLett.110.071102}%
  \BibitemOpen
  \bibfield  {author} {\bibinfo {author} {\bibfnamefont {U.}~\bibnamefont
  {Das}}\ and\ \bibinfo {author} {\bibfnamefont {B.}~\bibnamefont
  {Mukhopadhyay}},\ }\href {\doibase 10.1103/PhysRevLett.110.071102} {\bibfield
   {journal} {\bibinfo  {journal} {Phys. Rev. Lett.}\ }\textbf {\bibinfo
  {volume} {110}},\ \bibinfo {pages} {071102} (\bibinfo {year}
  {2013})}\BibitemShut {NoStop}%
\bibitem [{\citenamefont {Coelho}\ \emph {et~al.}(2014)\citenamefont {Coelho},
  \citenamefont {Marinho}, \citenamefont {Malheiro}, \citenamefont {Negreiros},
  \citenamefont {Rueda} \emph {et~al.}}]{Coelho:2013bba}%
  \BibitemOpen
  \bibfield  {author} {\bibinfo {author} {\bibfnamefont {J.}~\bibnamefont
  {Coelho}}, \bibinfo {author} {\bibfnamefont {R.}~\bibnamefont {Marinho}},
  \bibinfo {author} {\bibfnamefont {M.}~\bibnamefont {Malheiro}}, \bibinfo
  {author} {\bibfnamefont {R.}~\bibnamefont {Negreiros}}, \bibinfo {author}
  {\bibfnamefont {J.}~\bibnamefont {Rueda}},  \emph {et~al.},\ }\href {\doibase
  10.1088/0004-637X/794/1/86} {\bibfield  {journal} {\bibinfo  {journal}
  {Astrophys.J.}\ }\textbf {\bibinfo {volume} {794}},\ \bibinfo {pages} {86}
  (\bibinfo {year} {2014})},\ \Eprint {http://arxiv.org/abs/1306.4658}
  {arXiv:1306.4658 [astro-ph.SR]} \BibitemShut {NoStop}%
\bibitem [{\citenamefont {Chamel}\ \emph {et~al.}(2013)\citenamefont {Chamel},
  \citenamefont {Fantina},\ and\ \citenamefont {Davis}}]{Chamel:2013tfa}%
  \BibitemOpen
  \bibfield  {author} {\bibinfo {author} {\bibfnamefont {N.}~\bibnamefont
  {Chamel}}, \bibinfo {author} {\bibfnamefont {A.}~\bibnamefont {Fantina}}, \
  and\ \bibinfo {author} {\bibfnamefont {P.}~\bibnamefont {Davis}},\ }\href
  {\doibase 10.1103/PhysRevD.88.081301} {\bibfield  {journal} {\bibinfo
  {journal} {Phys.Rev.}\ }\textbf {\bibinfo {volume} {D88}},\ \bibinfo {pages}
  {081301} (\bibinfo {year} {2013})},\ \Eprint {http://arxiv.org/abs/1306.3444}
  {arXiv:1306.3444 [astro-ph.SR]} \BibitemShut {NoStop}%
\bibitem [{\citenamefont {Nityananda}\ and\ \citenamefont
  {Konar}(2015{\natexlab{a}})}]{PhysRevD.91.028301}%
  \BibitemOpen
  \bibfield  {author} {\bibinfo {author} {\bibfnamefont {R.}~\bibnamefont
  {Nityananda}}\ and\ \bibinfo {author} {\bibfnamefont {S.}~\bibnamefont
  {Konar}},\ }\href {\doibase 10.1103/PhysRevD.91.028301} {\bibfield  {journal}
  {\bibinfo  {journal} {Phys. Rev. D}\ }\textbf {\bibinfo {volume} {91}},\
  \bibinfo {pages} {028301} (\bibinfo {year} {2015}{\natexlab{a}})}\BibitemShut
  {NoStop}%
\bibitem [{\citenamefont {Das}\ and\ \citenamefont
  {Mukhopadhyay}(2014{\natexlab{a}})}]{Das:2014ssa}%
  \BibitemOpen
  \bibfield  {author} {\bibinfo {author} {\bibfnamefont {U.}~\bibnamefont
  {Das}}\ and\ \bibinfo {author} {\bibfnamefont {B.}~\bibnamefont
  {Mukhopadhyay}},\ }\href {\doibase 10.1088/1475-7516/2014/06/050} {\bibfield
  {journal} {\bibinfo  {journal} {JCAP}\ }\textbf {\bibinfo {volume} {1406}},\
  \bibinfo {pages} {050} (\bibinfo {year} {2014}{\natexlab{a}})},\ \Eprint
  {http://arxiv.org/abs/1404.7627} {arXiv:1404.7627 [astro-ph.SR]} \BibitemShut
  {NoStop}%
\bibitem [{\citenamefont {Das}\ and\ \citenamefont
  {Mukhopadhyay}(2014{\natexlab{b}})}]{Das:2013kga}%
  \BibitemOpen
  \bibfield  {author} {\bibinfo {author} {\bibfnamefont {U.}~\bibnamefont
  {Das}}\ and\ \bibinfo {author} {\bibfnamefont {B.}~\bibnamefont
  {Mukhopadhyay}},\ }\href {\doibase 10.1142/S0217732314500357} {\bibfield
  {journal} {\bibinfo  {journal} {Mod.Phys.Lett.}\ }\textbf {\bibinfo {volume}
  {A29}},\ \bibinfo {pages} {1450035} (\bibinfo {year} {2014}{\natexlab{b}})},\
  \Eprint {http://arxiv.org/abs/1304.3022} {arXiv:1304.3022 [astro-ph.SR]}
  \BibitemShut {NoStop}%
\bibitem [{\citenamefont {Das}\ and\ \citenamefont
  {Mukhopadhyay}(2015)}]{PhysRevD.91.028302}%
  \BibitemOpen
  \bibfield  {author} {\bibinfo {author} {\bibfnamefont {U.}~\bibnamefont
  {Das}}\ and\ \bibinfo {author} {\bibfnamefont {B.}~\bibnamefont
  {Mukhopadhyay}},\ }\href {\doibase 10.1103/PhysRevD.91.028302} {\bibfield
  {journal} {\bibinfo  {journal} {Phys. Rev. D}\ }\textbf {\bibinfo {volume}
  {91}},\ \bibinfo {pages} {028302} (\bibinfo {year} {2015})}\BibitemShut
  {NoStop}%
\bibitem [{\citenamefont {Nityananda}\ and\ \citenamefont
  {Konar}(2014)}]{Nityananda:2013yua}%
  \BibitemOpen
  \bibfield  {author} {\bibinfo {author} {\bibfnamefont {R.}~\bibnamefont
  {Nityananda}}\ and\ \bibinfo {author} {\bibfnamefont {S.}~\bibnamefont
  {Konar}},\ }\href {\doibase 10.1103/PhysRevD.91.029904,
  10.1103/PhysRevD.89.103017} {\bibfield  {journal} {\bibinfo  {journal}
  {Phys.Rev.}\ }\textbf {\bibinfo {volume} {D89}},\ \bibinfo {pages} {103017}
  (\bibinfo {year} {2014})},\ \Eprint {http://arxiv.org/abs/1306.1625}
  {arXiv:1306.1625 [astro-ph.SR]} \BibitemShut {NoStop}%
\bibitem [{\citenamefont {Nityananda}\ and\ \citenamefont
  {Konar}(2015{\natexlab{b}})}]{PhysRevD.91.029904}%
  \BibitemOpen
  \bibfield  {author} {\bibinfo {author} {\bibfnamefont {R.}~\bibnamefont
  {Nityananda}}\ and\ \bibinfo {author} {\bibfnamefont {S.}~\bibnamefont
  {Konar}},\ }\href {\doibase 10.1103/PhysRevD.91.029904} {\bibfield  {journal}
  {\bibinfo  {journal} {Phys. Rev. D}\ }\textbf {\bibinfo {volume} {91}},\
  \bibinfo {pages} {029904} (\bibinfo {year} {2015}{\natexlab{b}})}\BibitemShut
  {NoStop}%
\bibitem [{\citenamefont {Suh}\ and\ \citenamefont
  {Mathews}(2000)}]{Suh:1999tg}%
  \BibitemOpen
  \bibfield  {author} {\bibinfo {author} {\bibfnamefont {I.-S.}\ \bibnamefont
  {Suh}}\ and\ \bibinfo {author} {\bibfnamefont {G.}~\bibnamefont {Mathews}},\
  }\href {\doibase 10.1086/308403} {\bibfield  {journal} {\bibinfo  {journal}
  {Astrophys.J.}\ }\textbf {\bibinfo {volume} {530}},\ \bibinfo {pages} {949}
  (\bibinfo {year} {2000})},\ \Eprint {http://arxiv.org/abs/astro-ph/9906239}
  {arXiv:astro-ph/9906239 [astro-ph]} \BibitemShut {NoStop}%
\bibitem [{\citenamefont {Bowers}\ and\ \citenamefont
  {Liang}(1974)}]{Bowers:1974}%
  \BibitemOpen
  \bibfield  {author} {\bibinfo {author} {\bibfnamefont {R.~L.}\ \bibnamefont
  {Bowers}}\ and\ \bibinfo {author} {\bibfnamefont {E.~P.~T.}\ \bibnamefont
  {Liang}},\ }\href {\doibase 10.1086/152760} {\bibfield  {journal} {\bibinfo
  {journal} {Astrophys.J.}\ }\textbf {\bibinfo {volume} {188}},\ \bibinfo
  {pages} {657} (\bibinfo {year} {1974})}\BibitemShut {NoStop}%
\bibitem [{\citenamefont {Das}\ and\ \citenamefont
  {Mukhopadhyay}(2014{\natexlab{c}})}]{Das:2014owa}%
  \BibitemOpen
  \bibfield  {author} {\bibinfo {author} {\bibfnamefont {U.}~\bibnamefont
  {Das}}\ and\ \bibinfo {author} {\bibfnamefont {B.}~\bibnamefont
  {Mukhopadhyay}},\ }\href@noop {} {\  (\bibinfo {year}
  {2014}{\natexlab{c}})},\ \Eprint {http://arxiv.org/abs/1411.5367}
  {arXiv:1411.5367 [astro-ph.SR]} \BibitemShut {NoStop}%
\bibitem [{\citenamefont {Bocquet}\ \emph {et~al.}(1995)\citenamefont
  {Bocquet}, \citenamefont {Bonazzola}, \citenamefont {Gourgoulhon},\ and\
  \citenamefont {Novak}}]{Bocquet:1995je}%
  \BibitemOpen
  \bibfield  {author} {\bibinfo {author} {\bibfnamefont {M.}~\bibnamefont
  {Bocquet}}, \bibinfo {author} {\bibfnamefont {S.}~\bibnamefont {Bonazzola}},
  \bibinfo {author} {\bibfnamefont {E.}~\bibnamefont {Gourgoulhon}}, \ and\
  \bibinfo {author} {\bibfnamefont {J.}~\bibnamefont {Novak}},\ }\href@noop {}
  {\bibfield  {journal} {\bibinfo  {journal} {Astron.Astrophys.}\ }\textbf
  {\bibinfo {volume} {301}},\ \bibinfo {pages} {757} (\bibinfo {year}
  {1995})},\ \Eprint {http://arxiv.org/abs/gr-qc/9503044} {arXiv:gr-qc/9503044
  [gr-qc]} \BibitemShut {NoStop}%
\bibitem [{\citenamefont {Bucciantini}\ and\ \citenamefont
  {Del~Zanna}(2011)}]{Bucciantini:2010ax}%
  \BibitemOpen
  \bibfield  {author} {\bibinfo {author} {\bibfnamefont {N.}~\bibnamefont
  {Bucciantini}}\ and\ \bibinfo {author} {\bibfnamefont {L.}~\bibnamefont
  {Del~Zanna}},\ }\href {\doibase 10.1051/0004-6361/201015945} {\bibfield
  {journal} {\bibinfo  {journal} {Astron.Astrophys.}\ }\textbf {\bibinfo
  {volume} {528}},\ \bibinfo {pages} {A101} (\bibinfo {year} {2011})},\ \Eprint
  {http://arxiv.org/abs/1010.3532} {arXiv:1010.3532 [astro-ph.IM]} \BibitemShut
  {NoStop}%
\bibitem [{\citenamefont {Camenzind}(2007)}]{Max}%
  \BibitemOpen
  \bibfield  {author} {\bibinfo {author} {\bibfnamefont {M.}~\bibnamefont
  {Camenzind}},\ }\href@noop {} {\emph {\bibinfo {title} {Compact Objects in
  Astrophysics: White Dwarfs, Neutron Stars and Black Holes}}},\ \bibinfo
  {number} {137-186}\ (\bibinfo  {publisher} {Spinger},\ \bibinfo {year}
  {2007})\BibitemShut {NoStop}%
\bibitem [{\citenamefont {Misner}\ \emph {et~al.}(1973)\citenamefont {Misner},
  \citenamefont {Thorne},\ and\ \citenamefont {Wheeler}}]{MTW}%
  \BibitemOpen
  \bibfield  {author} {\bibinfo {author} {\bibfnamefont {C.}~\bibnamefont
  {Misner}}, \bibinfo {author} {\bibfnamefont {K.}~\bibnamefont {Thorne}}, \
  and\ \bibinfo {author} {\bibfnamefont {J.}~\bibnamefont {Wheeler}},\
  }\href@noop {} {\emph {\bibinfo {title} {Gravitation}}},\ \bibinfo {number}
  {557-590}\ (\bibinfo  {publisher} {W. H. Freeman and Company: San
  Francisco},\ \bibinfo {year} {1973})\BibitemShut {NoStop}%
\bibitem [{\citenamefont {Ginzberg}(1964)}]{Ginzberg:1964}%
  \BibitemOpen
  \bibfield  {author} {\bibinfo {author} {\bibfnamefont {V.}~\bibnamefont
  {Ginzberg}},\ }\href@noop {} {\bibfield  {journal} {\bibinfo  {journal}
  {Sov.Phys.Dokl.}\ }\textbf {\bibinfo {volume} {9}},\ \bibinfo {pages} {239}
  (\bibinfo {year} {1964})}\BibitemShut {NoStop}%
\bibitem [{\citenamefont {Thorne}(1964)}]{thorne1964resistance}%
  \BibitemOpen
  \bibfield  {author} {\bibinfo {author} {\bibfnamefont {K.~S.}\ \bibnamefont
  {Thorne}},\ }in\ \href@noop {} {\emph {\bibinfo {booktitle} {Quasars and High
  Energy Astronomy; Including the Proceedings of the Second Texas Symposium on
  Relativistic Astrophysics}}}\ (\bibinfo {year} {1964})\BibitemShut {NoStop}%
\bibitem [{\citenamefont {Woltjer}(1964)}]{woltjer1964x}%
  \BibitemOpen
  \bibfield  {author} {\bibinfo {author} {\bibfnamefont {L.}~\bibnamefont
  {Woltjer}},\ }\href@noop {} {\bibfield  {journal} {\bibinfo  {journal} {The
  Astrophysical Journal}\ }\textbf {\bibinfo {volume} {140}},\ \bibinfo {pages}
  {1309} (\bibinfo {year} {1964})}\BibitemShut {NoStop}%
\bibitem [{\citenamefont {Jackson}(1999)}]{Jackson}%
  \BibitemOpen
  \bibfield  {author} {\bibinfo {author} {\bibfnamefont {J.}~\bibnamefont
  {Jackson}},\ }\href@noop {} {\emph {\bibinfo {title} {Classical
  Electrodynamics}}},\ \bibinfo {number} {198-199}\ (\bibinfo  {publisher}
  {John Wiley $\& $ Sons, Inc.},\ \bibinfo {year} {1999})\BibitemShut {NoStop}%
\bibitem [{\citenamefont {Wang}\ and\ \citenamefont {Guo}(1989)}]{SpeFunc}%
  \BibitemOpen
  \bibfield  {author} {\bibinfo {author} {\bibfnamefont {Z.}~\bibnamefont
  {Wang}}\ and\ \bibinfo {author} {\bibfnamefont {D.}~\bibnamefont {Guo}},\
  }\href@noop {} {\emph {\bibinfo {title} {Special Functions}}},\ \bibinfo
  {number} {8-14}\ (\bibinfo  {publisher} {World Scientific},\ \bibinfo {year}
  {1989})\BibitemShut {NoStop}%
\bibitem [{\citenamefont {Bera}\ and\ \citenamefont
  {Bhattacharya}(2014)}]{Bera:2014wja}%
  \BibitemOpen
  \bibfield  {author} {\bibinfo {author} {\bibfnamefont {P.}~\bibnamefont
  {Bera}}\ and\ \bibinfo {author} {\bibfnamefont {D.}~\bibnamefont
  {Bhattacharya}},\ }\href {\doibase 10.1093/mnras/stu2014} {\bibfield
  {journal} {\bibinfo  {journal} {Mon.Not.Roy.Astron.Soc.}\ }\textbf {\bibinfo
  {volume} {445}},\ \bibinfo {pages} {3951} (\bibinfo {year} {2014})},\ \Eprint
  {http://arxiv.org/abs/1405.2282} {arXiv:1405.2282 [astro-ph.SR]} \BibitemShut
  {NoStop}%
\end{thebibliography}%

\end{document}